\documentclass[12pt]{article}
\PassOptionsToPackage{hyphens}{url}
\usepackage[pagebackref=true,colorlinks]{hyperref}
\usepackage{amsmath,amssymb,amsthm,mathtools,amsrefs,mathrsfs,marvosym}
\usepackage[utf8]{inputenc}
\usepackage[T1]{fontenc}
\usepackage{pifont}% http://ctan.org/pkg/pifont
\usepackage{mathpazo}
\usepackage{authblk} % add possibly `sc` and `osf` options
\usepackage{eulervm}
\usepackage{csquotes}
\usepackage{bbm}
\usepackage{lscape}%puts selected page in landscape mode
\usepackage{etex} %allows xRightarrow
\usepackage[table]{xcolor}%This is used to add colors to rows and columns in arrays (which I use for matrices) WARNING: Must load before tikz packages (i believe arrows.meta uses xcolor so there is a clash). or another option is to add "table" as a global option to the document class, ex. \documentclass[12pt,table]{article}
\usepackage{tikz}
\usetikzlibrary{arrows.meta}
\usetikzlibrary{decorations.pathmorphing,shapes}
\usetikzlibrary{calc}
\usepackage{comment}
\usepackage{adjustbox}
\usepackage{scalerel,stackengine}%Used for reallywidehat
\usepackage{stmaryrd}
\usepackage{fancyhdr}
\usepackage{soul}
\usepackage{dsfont}%this is for mathbb{1} but you write mathds{1}
\usepackage[normalem]{ulem}
\usepackage{longtable}
\usepackage[bottom]{footmisc}%this forces footnotes to appear on the bottom of page
\usepackage{graphicx}
\usepackage{caption}
\usepackage{subcaption}
\usepackage{pictexwd, dcpic}
\usepackage{float}
\usepackage{enumerate}
\usepackage[all]{xy} %This is to make higher categorical stuff
\usetikzlibrary{circuits.ee.IEC}
\hypersetup{
	colorlinks=true,%set true if you want colored links
    linktoc=all,%set to all if you want both sections and subsections linked
    linkcolor=violet,  %choose some color if you want links to stand out
    citecolor=green!70!black,
    }
\usepackage{tocbasic}
\DeclareTOCStyleEntry[
  beforeskip=-0.3em plus 1pt,% default is 1em plus 1pt
  pagenumberformat=\textbf
]{tocline}{section}
\makeatletter
\newcommand*{\shifttext}[2]{%
  \settowidth{\@tempdima}{#2}%
  \makebox[\@tempdima]{\hspace*{#1}#2}%
}
\makeatother
\makeatletter
\renewcommand*\env@matrix[1][\arraystretch]{%
  \edef\arraystretch{#1}%
  \hskip -\arraycolsep
  \let\@ifnextchar\new@ifnextchar
  \array{*\c@MaxMatrixCols c}}
\makeatother

\stackMath
\newcommand\reallywidehat[1]{%
\savestack{\tmpbox}{\stretchto{%
  \scaleto{%
    \scalerel*[\widthof{\ensuremath{#1}}]{\kern.1pt\mathchar"0362\kern.1pt}%
    {\rule{0ex}{\textheight}}%WIDTH-LIMITED CIRCUMFLEX
  }{\textheight}% 
}{2.4ex}}%
\stackon[-6.9pt]{#1}{\tmpbox}%
}
\makeatletter
\pgfmathdeclarefunction{erf}{1}{%
  \begingroup
    \pgfmathparse{#1 > 0 ? 1 : -1}%
    \edef\sign{\pgfmathresult}%
    \pgfmathparse{abs(#1)}%
    \edef\x{\pgfmathresult}%
    \pgfmathparse{1/(1+0.3275911*\x)}%
    \edef\t{\pgfmathresult}%
    \pgfmathparse{%
      1 - (((((1.061405429*\t -1.453152027)*\t) + 1.421413741)*\t 
      -0.284496736)*\t + 0.254829592)*\t*exp(-(\x*\x))}%
    \edef\y{\pgfmathresult}%
    \pgfmathparse{(\sign)*\y}%
    \pgfmath@smuggleone\pgfmathresult%
  \endgroup
}
\makeatother
\theoremstyle{theorem}
\newtheorem{theorem}[equation]{Theorem}
\newtheorem{lemma}[equation]{Lemma}
\newtheorem{proposition}[equation]{Proposition}
\newtheorem{corollary}[equation]{Corollary}
\theoremstyle{definition}
\newtheorem{definition}[equation]{Definition}
\newtheorem{construction}[equation]{Construction}
\newtheorem{question}[equation]{Question}

\newtheorem{problem}[equation]{Problem}
\newtheorem{example}[equation]{Example}
\newtheorem{exercise}[equation]{Exercise}
\newtheorem*{answer}{Answer}
\newtheorem*{solution}{Solution}
\newtheorem{remark}[equation]{Remark}

\newtheorem{notation}[equation]{Notation}
\newtheorem{noterm}[equation]{Notation and Terminology}

\newcommand\define[1]{\emph{\textbf{#1}}}%italicize and bold-face %this seems like a good alternative

\numberwithin{equation}{section}
\newcommand{\be}{\begin{equation}}
\newcommand{\ee}{\end{equation}}
\def\ba{\begin{align}} %previously this was ``array''
\def\ea{\end{align}}
\newcommand{\bea}{\begin{eqnarray}}
\newcommand{\eea}{\end{eqnarray}}
\newcommand{\bx}{\begin{example}}
\newcommand{\ex}{\end{example}}
\newcommand{\bex}{\begin{exercise}}
\newcommand{\eex}{\end{exercise}}
\newcommand{\ban}{\begin{answer}}
\newcommand{\ean}{\end{answer}}
\newcommand{\bt}{\begin{theorem}}
\newcommand{\et}{\end{theorem}}
\newcommand{\bc}{\begin{corollary}}
\newcommand{\ec}{\end{corollary}}
\newcommand{\blem}{\begin{lemma}}
\newcommand{\elem}{\end{lemma}}
\newcommand{\bp}{\begin{problem}}
\newcommand{\ep}{\end{problem}}
\newcommand{\bn}{\begin{proposition}}
\newcommand{\en}{\end{proposition}}
\newcommand{\bd}{\begin{definition}}
\newcommand{\ed}{\end{definition}}
\newcommand{\bcon}{\begin{construction}}
\newcommand{\econ}{\end{construction}}
\newcommand{\bq}{\begin{question}}
\newcommand{\eq}{\end{question}}
\newcommand{\bprf}{\begin{proof}}
\newcommand{\eprf}{\end{proof}}
\newcommand{\br}{\begin{remark}}
\newcommand{\er}{\end{remark}}
\newcommand{\bs}{\begin{solution}}
\newcommand{\es}{\end{solution}}
\newcommand{\beqs}{\begin{eqnarray}}
\newcommand{\eeqs}{\end{eqnarray}}
\newcommand{\bnt}{\begin{noterm}}
\newcommand{\ent}{\end{noterm}}
\newcommand{\bnot}{\begin{notation}}
\newcommand{\enot}{\end{notation}}

\newcommand{\lra}{\longrightarrow}

\newcommand{\Tr}{{\rm Tr} }

\def\R{{{\mathbb R}}}

\def\X{\mathfrak{X}}

\newcommand{\stoch}{\;\xy0;/r.25pc/:(-3,0)*{}="1";(3,0)*{}="2";{\ar@{~>}"1";"2"|(1.06){\hole}};\endxy\!}
\newcounter{sarrow}

\newcounter{sqarrow}

\DeclareFontFamily{OT1}{pzc}{}
\DeclareFontShape{OT1}{pzc}{m}{it}{ <-> s*[1.2] pzcmi7t }{}
\DeclareMathAlphabet{\mathpzc}{OT1}{pzc}{m}{it}

\newcommand{\ds}{\displaystyle}
\newcommand{\ben}{\renewcommand{\theenumi}{\alph{enumi}} 
\renewcommand{\labelenumi}{(\theenumi)}\begin{enumerate}}
\newcommand{\een}{\end{enumerate}}
\newcommand*{\matminus}{%
  \leavevmode
  \hphantom{0}%
  \llap{%
    \settowidth{\dimen0 }{$0$}%
    \resizebox{1.1\dimen0 }{\height}{$-$}%
  }%
}

\title{A diagrammatic formulation of local realism}
\author[$\spadesuit$]{James Fullwood}
\affil[$\spadesuit$]{School of Mathematics and Statistics, Hainan University, Haikou, Hainan, 570228, China}
\date{}                     %% if you don't need date to appear
\begin{document}
\emergencystretch 2em

\maketitle

\begin{abstract}
Given two parties performing experiments in separate laboratories, we provide a diagrammatic formulation of what it means for the joint statistics of their experiments to satisfy local realism. In particular, we show that the principles of locality and realism are both captured by a single commutative diagram in the category of probability-preserving maps between finite probability spaces, and we also show that an assumption of such a diagrammatic formulation of local realism implies the standard CHSH inequality associated with dichotomic random variables. As quantum theory is known not to satisfy local realism, our formulation of local realism in terms of commutative diagrams provides yet another way in which the notion of non-commutativity plays a fundamental role in quantum theory. We note that we do not assume any prior knowledge of category theory or quantum theory, as this work is intended for philosophers, mathematicians and physicists alike.

%As this work only assumes basic knowledge of classical probability theory on finite sets, it is intended for philosophers, mathematicians and physicists alike. In particular, no prior knowledge of categories or quantum theory is assumed in this work.
\end{abstract}

\vspace{-7mm}
\tableofcontents

\newpage

%%%%%%%%%%%%%
\section{Introduction}
%%%%%%%%%%%%%

In one of the most cited papers in all of physics \cite{EPR}, Einstein, Podolsky and Rosen (EPR) formulated an argument for why quantum theory cannot be a complete description of reality based upon the physical principles of locality and realism. While locality is the principle that a physical system cannot causally influence events outside of its future lightcone, realism is the principle that properties of physical systems which are revealed by measurement are intrinsic, and exist whether or not we actually choose to measure those properties. The principles of locality and realism are often combined to form the principle of \emph{local realism}, and the viewpoint espoused by EPR was that local realism should be satisfied by any fundamental theory of physics. As quantum theory was known at the time to exhibit characteristics which violate local realism, EPR argued that quantum theory was then an incomplete description of reality, and merely an intermediary step on our way to a more fundamental theory which was locally real. 

After publication of the EPR paper in 1935, the notion of a realistic theory consistent with the predictions of quantum theory eventually came to be synonymous with the notion of a "hidden variables" theory, in which the probabilistic nature of quantum theory is viewed as being due to unseen sources which are in fact deterministic. For example, in 1952 Bohm formulated a theory of the pilot wave \cite{Bohm52}, which unbeknownst to Bohm, was a theory which had first been introduced by de Broglie at the Solvay conference in 1927~\cite{deBroglie_1926}, but later abandoned due to criticism. In the de Broglie-Bohm theory, particles such as an electron passing through a double-slit are guided by an undetectable wave, thus resulting in classical trajectories for quantum particles. While the de Broglie-Bohm theory was successful in getting rid of the inherent randomness of quantum theory, it was manifestly non-local, and hence still not a theory as envisioned by EPR.

Almost 30 years after the EPR paper was published, John Bell proved results which came to be known as \emph{Bell's Theorem} \cite{Bell64}, which unequivocally established the fact that any theory which satisfies local realism cannot be consistent with the predictions of quantum theory.  In the proof of Bell's Theorem, Bell considered a situation in which two parties in remote locations perform an experiment, where a single run of the experiment consists of each party randomly making a choice to perform one out of a list of two measurements. In the case that the two parties label their measurement outcomes by $\pm 1$, Bell was able to show that an assumption of local realism leads to a constraint on the joint measurement statistics of the experiment known today as Bell's inequality. Bell then showed that quantum theory predicts a violation of this inequality for a particular situation in which the two parties are performing spin measurements on an entangled pair of electrons, from which it follows that any theory consistent with quantum theory cannot be locally real. From the 1970s to the 1990s many physicists worked toward a realization of an experimental scenario whose statistics would violate Bell's inequality, and in 2022, Clauser, Aspect and Zeilinger were awarded the Nobel Prize in physics for their work leading to an experimental verification of the failure of local realism~\cite{Clauser_1969,Clauser_1972,Aspect_1981,Greenberger_1989,Bouwmeester_1998}.

Although the physical principles of locality and realism are intuitively clear, translating them into precise mathematics in the context of quantum measurements is not so straightforward. In particular, anyone who takes the time to search the literature will find that there isn't a general consensus regarding a mathematically precise definition of local realism. As for the principle of realism, the situation is succinctly captured in the words of Asher Peres, who writes that realism has ``at least as many definitions as there are authors''~\cite{Peres88}. And this is not at all surprising, since even in pure mathematics it often takes a significant amount of time before a precise, working definition of a fundamental concept comes to light. For example, while the intuitively clear concept of continuity is fundamental to integral calculus as first established by Newton and Leibniz in the 17th century, it wasn't until Cauchy's work in the early 19th century that a rigorous and mathematically precise notion of continuity was first formulated.    

%And this is not at all surprising, since even in pure mathematics it may decades before a simple, working definition of a fundamental concept comes to light. For example, \JF{Perhaps use the example of the concept of continuity instead, or maybe entropy} although it was known since the 1830s that any continuous function on a closed and bounded subset of $\R^n$ must necessarily attain a maximum and minimum value, it was unclear for a period of time how to generalize this result---which came to be known as the \emph{Extreme Value Theorem}--- to general topological spaces. In particular, while the notion of being closed makes sense for any topological space, there is no notion of a subset being bounded for general topological spaces, and so for a significant period of time it was not clear how to formulate the Extreme Value Theorem in a general topological setting. As such, it wasn't until the establishment of the Heine-Borel Theorem in the late 1890s that the simple and elegant definition of a compact set emerged, thus providing the key ingredient for a generalization of the Extreme Value Theorem. 

%And this is not at all surprising, as it is often the case that it takes a significant amount of time for fundamental concepts in physics to take on a solid mathematical form. For example, while the concept which came to be known as entropy was first introduced by Clausius in 1854, the modern mathematical definition as formulated by Gibbs and Boltzman did not appear until 1877.

In this work, we use probability-preserving maps between finite probability spaces to formulate a simple and precise mathematical definition of local realism. Moreover, by viewing finite probability spaces and probability-preserving maps between them as the objects and morphisms in what mathematicians refer to as a \emph{category}, we are able to formulate the principle of local realism as a certain commutative diagram in this category. We then show that our diagrammatic formulation of local realism is consistent with Bell's Theorem,  as we show that under an assumption of our diagrammatic formulation of local realism one may derive the standard CHSH inequality associated with dichotomic random variables (see Theorem~\ref{TMX687}). The CHSH inequality is a simplification of Bell's original equality first derived by Clauser, Horne, Shimony and Holt in Ref.~\cite{Cl69}, which was crucial for making Bell's Theorem amenable to experiment. In a similar vein, our diagrammatic formulation of local realism may be viewed as a simplification of Bell's original distillation of the ideas of EPR, and the use of commutative diagrams makes the notion of local realism applicable in a broader context than that of Bell's Theorem.

We note that although we use the language of categories and commutative diagrams for our formulation of local realism, we emphasize that no prior knowledge of categories is assumed in this work, as we only assume a basic knowledge of probability theory on finite sets. In particular, the use of categorical diagrams is simply employed as a way in which to organize the data defining typical measurement scenarios as those considered by Bell, in order to elucidate their structural properties. Moreover, we also do not assume any prior knowledge of quantum theory, as this work is intended for philosophers, mathematicians and physicists alike.

This paper is organized as follows. In Section~\ref{BXRXC}, we briefly state what categories and commutative diagrams are without making any formal definitions, while also motivating the use of categorical language in the paper. In Section~\ref{FPSX}, we give the formal mathematical definitions necessary for our diagrammatic formulation of local realism, such as that of finite probability space, probability-preserving maps between probability spaces and that of a random variable and its pullback under a probability-preserving map. In Section~\ref{BSXTMXS47} we recall the type of measurement scenario considered by Bell in \cite{Bell64}, which we refer to as a \emph{Bell scenario}. We then formally define what we refer to as a \emph{theory of measurement} in the context of Bell scenarios, setting the stage for our diagrammatic formulation of local realism. In Section~\ref{RLX87}, we provide our formal diagrammatic definition of what it means for the joint statistics associated with a Bell scenario to satisfy local realism. While the principle of local realism is captured by a single diagram, we show how the individual principles of locality and realism may also be formulated diagrammatically. We also show how the notion of a hidden variables theory is made precise by our diagrammatic formalism. In Section~\ref{CHSH}, we derive the standard CHSH inequality under an assumption of our diagrammatic formulation of local realism, and we give an example showing how quantum theory predicts the failure of local realism. For those not familiar with quantum theory, in the appendix we give an explicit example of how joint probabilities associated with a Bell scenario are computed in quantum theory.

\noindent \emph{Acknowledgements}. We thank Tobias Fritz and Arthur J. Parzygnat for many useful discussions, and we further thank Arthur J. Parzygnat for pointing out that locality and realism may be separately formulated in terms of the commutative diagrams which appear in Figure~1. 

%%%%%%%%%%%%%%%%%%%%%%%%%%%%
\section{A brief remark on categories}\label{BXRXC}
%%%%%%%%%%%%%%%%%%%%%%%%%%%%

A category is a mathematical structure which was introduced to organize mathematical discourse \cite{Ma98}, so that we may identify recurring structures in seemingly disparate fields of mathematics. In particular, most mathematicians work in a specialized field, where they consider objects of a certain type together with a certain class mappings between those objects. Category theory is then a way to make this observation more formal, so that a category $\mathscr{C}$ is defined to be a certain collection of mathematical entities which we refer to as the  \emph{objects} of $\mathscr{C}$, together with a certain collection of mappings between the objects, which we refer to as the \emph{morphisms} of $\mathscr{C}$. For example, a topologist will often concern themselves with continuous mappings between manifolds, so that from a category-theoretic perspective, a topologist works in the category $\bold{Top}$, whose objects are manifolds and whose morphisms are continuous mappings between manifolds. A combinatorialist on the other hand may often work in the category $\bold{Fin}$, whose objects are finite sets and whose morphisms are bijective functions between finite sets.

 If $A$ and $B$ are objects of a category $\mathscr{C}$ and $f$ is a morphism corresponding to a mapping from $A$ to $B$, then the morphism $f$ is typically denoted by an arrow $A\overset{f}\to B$. Moreover, if $B\overset{g}\to C$ is another morphism in $\mathscr{C}$, then one of the axioms in the formal definition of a category ensures that we can \emph{compose} the morphisms $f$ and $g$ to obtain a new morphism $A\overset{g\circ f}\lra C$. As such, when one wishes to introduce a category into their work, they often do so by simply stating what the objects and the morphisms are, together with a rule for how the morphisms compose. 
 
 As category theory places an emphasis on structure, one of the primary philosophies of a category theorist is to phrase fundamental concepts in terms of \emph{diagrams} in a category. A diagram in a category $\mathscr{C}$ is simply a directed graph whose nodes are labeled by objects in $\mathscr{C}$ and whose directed edges are labeled by morphisms in $\mathscr{C}$, and the collection of all diagrams in a category $\mathscr{C}$ will be denoted by $\mathfrak{Diag}(\mathscr{C})$. For example, a diagram may look like
 
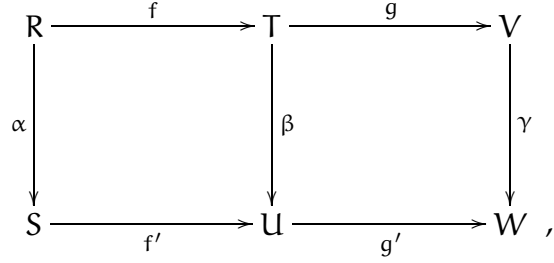
\[
\xy0;/r.25pc/:
(0,0)*+{U}="0";
(-30,0)*+{S}="1";
(-30,25)*+{R}="3";
(0,25)*+{T}="4";
(30,0)*+{W}="8";
(30,25)*+{V}="9";
{\ar"1";"0"_{f'}};
{\ar"0";"8"_{g'}};
{\ar"4";"0"^{\beta}};
{\ar"3";"4"^{f}};
{\ar"9";"8"^{\gamma}};
{\ar"4";"9"^{g}};
{\ar"3";"1"_{\alpha}}; 
\endxy \, ,
 \]
 and such a diagram is said to be \emph{commutative} if and only if every path between two fixed nodes yields the same result. For example, the diagram above is commutative if and only if
 \[
 \beta\circ f = f'\circ \alpha\, , \quad \gamma \circ g=g'\circ \beta\, , \quad \text{and} \quad \gamma\circ g\circ f=g'\circ f'\circ \alpha\, .
 \]
 The notion of commutative diagram generalizes the notion of commutativity in algebra, where performing operations in two different ways may yield the same result, as reflected in the equation $ab=ba$. 

When fundamental concepts are phrased in terms of diagrams, it is often the case that such a diagrammatical formulation makes the concepts more structurally clear and precise. Moreover, when using diagrams to formulate concepts, one can then also migrate such concepts from one category to another by relabeling the edges and nodes of the associated diagram by the objects and morphisms of a \emph{different} category, which is a process made mathematically precise by the notion of a \emph{functor}. While we won't need functors for this work, we will adopt the category theorist's philosophy by phrasing the physical principle of local realism in terms of a certain diagram in the category of probability-preserving maps between finite probability spaces (which we define in the next section). By doing so, it is our hope that not only is the principle of local realism made mathematically precise, but that it is also made simple, clear, and intuitive. Moreover, giving a diagrammatic formulation of local realism sets the stage for the concept of local realism to be defined in more general probabilistic settings, such as Markov categories~\cite{Fr20} and general probabilistic theories (GPTs)~\cite{Segal_47,Plavala_2021}

%%%%%%%%%%%%%%%%%%%%%%%%%%%%%%%%%%%%
\section{Finite probability spaces and random variables}\label{FPSX}
%%%%%%%%%%%%%%%%%%%%%%%%%%%%%%%%%%%%

In this section we recall the basic mathematical definitions needed for the formulation of all mathematical statements in this work. 

\bd
A \define{finite probability space} consists of a pair $(\Omega_X,\mu)$, where $\Omega_X$ is a finite set and $\mu:\Omega_X\to [0,1]$ is a function such that
\[
\sum_{x\in X}\mu(x)=1\, .
\]
\ed

\br
In the context of this work, the underlying set $\Omega_X$ of a finite probability space $(\Omega_X,\mu)$ is thought of as a set whose elements label all possible outcomes of some experiment or random process labeled by $X$, while the function $\mu$ provides the theoretical likelihoods associated with every outcome $x\in \Omega_X$.
\er

\bd
The category $\bold{FinProb}$ is the category whose objects consist of finite probability spaces, and a morphism from $(\Omega_X,\mu)$ to $(\Omega_Y,\nu)$ consists of a function $f:\Omega_X\to \Omega_Y$ such that 
\be \label{PSWXS73XS}
\nu(y)=\sum_{x\in f^{-1}(y)}\mu(x) \qquad \forall y\in \Omega_Y\, .
\ee
Such a morphism will then be denoted by $(\Omega_X,\mu)\overset{f}\lra (\Omega_Y,\nu)$, and in such a case, the probability space $(\Omega_X,\mu)$ is said to be an \define{extension} of the probability space $(\Omega_Y,\nu)$. Given morphisms $(\Omega_X,\mu)\overset{f}\lra (\Omega_Y,\nu)$ and $(\Omega_Y,\nu)\overset{g}\lra (\Omega_Z,\lambda)$, their associated composition is the morphism $(\Omega_X,\mu)\overset{h}\lra (\Omega_Z,\lambda)$ with $h=g\circ f$.
\ed

\bd
Given finite probability spaces $(\Omega_X,\mu)$ and $(\Omega_Y,\nu)$, a finite probability space $(\Omega_X\times \Omega_Y,\vartheta)$ is said to be a \define{joint probability space} associated with  $(\Omega_X,\mu)$ and $(\Omega_Y,\nu)$ if and only if 
\be \label{MRGNXSL87}
\mu(x)=\sum_{y\in \Omega_Y}\vartheta(x,y) \quad \text{and} \quad \nu(y)=\sum_{x\in \Omega_X}\vartheta(x,y)  \qquad \forall\,\, x\in \Omega_X\,\, y\in \Omega_Y\, .
\ee
Conversely, given a probability space $(\Omega_X\times \Omega_Y,\vartheta)$, the probability spaces $(\Omega_X,\mu)$ and $(\Omega_Y,\nu)$ with $\mu$ and $\nu$ given by \eqref{MRGNXSL87} are then said to be the \define{marginal distributions} associated with $(\Omega_X\times \Omega_Y,\vartheta)$. In such a case, the canonical projections $\pi_X:\Omega_X\times \Omega_Y\to \Omega_X$ and $\pi_Y:\Omega_X\times \Omega_Y\to \Omega_Y$ given by $\pi_X(x,y)=x$ and $\pi_Y(x,y)=y$ then define morphisms 
\[
(\Omega_X\times \Omega_Y,\vartheta)\overset{\pi_X}\lra (\Omega_X,\mu) \quad \text{and} \quad (\Omega_X\times \Omega_Y,\vartheta)\overset{\pi_Y}\lra (\Omega_Y,\nu)
\]
in $\bold{FinProb}$.
\ed

\bd
Given finite probability spaces $(\Omega_X,\mu)$ and $(\Omega_Y,\nu)$, the associated \define{product distribution} is the joint probability space $(\Omega_X\times \Omega_Y,\vartheta)$ given by
\[
\vartheta(x,y)=p(x)q(y) \qquad \forall x\in \Omega_X\,\, y\in \Omega_Y\, .
\]
Such a product distribution will be denoted by $(\Omega_X\times \Omega_Y,\mu\times \nu)$.
\ed

\bd
A \define{random variable} on a finite probability space $(\Omega_X,\mu)$ consists of a real-valued function $\mathfrak{X}:\Omega_X\to \R$. The set of all random variables on $(\Omega_X,\mu)$ will be denoted by $\bold{RV}(\Omega_X,\mu)$, and the \define{expected value} of a random variable $\X\in \bold{RV}(\Omega_X,\mu)$ is the element $\bold{E}(\X)\in \R$ given by
\[
\bold{E}(\X)=\sum_{x\in \Omega_X}\X(x)\mu(x)\, .
\]
Given a morphism $(\Omega_W,\xi)\overset{h}\lra (\Omega_X,\mu)$ in $\bold{FinProb}$ and a random variable $\X\in \bold{RV}(\Omega_X,\mu)$, then the \define{pullback} of $\X$ with respect to $h$ is the random variable $h^*(\X)\in \bold{RV}(\Omega_W,\xi)$ given by 
\[
h^*(\X)(w)=(\X\circ h)(w) \qquad \forall \, w\in \Omega_W\, .
\]
\ed

\br
When the elements of a finite probability space are viewed as the outcomes of an experiment or random process, the notion of a random variable then allows one to associate real numbers with the outcomes, so that one may generate real-valued data from the experiment or random process. Once outcomes are mapped to data, then the outcomes become amenable to the plethora of tools developed for data-analysis. 
\er

The next proposition shows how the pullback of random variables preserves expected values, which will be crucial for Theorem~\ref{TMX687}.

\bn[Preservation of expected value under pullback] \label{EXPXT17XPB}
Let $(\Omega_W,\xi)\overset{h}\lra (\Omega_X,\mu)$ be a morphism in $\bold{FinProb}$, and let $\X\in \bold{RV}(\Omega_X,\mu)$. Then $\bold{E}(h^*(\X))=\bold{E}(\X)$.
\en

\bprf
Indeed,
\begin{align*}
\bold{E}(h^*(\X))&=\ds \sum_{w\in \Omega_W}h^{*}(\X)(w)\xi(w)=\ds \sum_{w\in \Omega_W}(\X\circ h)(w)\xi(w)=\ds \sum_{w\in \Omega_W}\X(h(w))\xi(w) \\
&=\ds \sum_{x\in \Omega_X}\left(\sum_{w\in h^{-1}(x)}\X(h(w))\xi(w)\right)=\ds \sum_{x\in \Omega_X}\left(\sum_{w\in h^{-1}(x)}\X(x)\xi(w)\right) \\
&=\ds \sum_{x\in \Omega_X}\X(x)\left(\sum_{w\in h^{-1}(x)}\xi(w)\right)=\ds \sum_{x\in \Omega_X}\X(x)p(x) \\
&=\bold{E}(\X)\, ,
\end{align*}
as desired.
\eprf

%%%%%%%%%%%%%%%%%%%%%%%%%%%%%%%%%%%%%%%
\section{Bell scenarios and theories of measurement} \label{BSXTMXS47}
%%%%%%%%%%%%%%%%%%%%%%%%%%%%%%%%%%%%%%%

A \emph{Bell scenario} consists of an experiment where Alice and Bob are in two separate laboratories, and a single run of the experiment consists of the following protocol:

\begin{itemize}
\item
A physical system in a fixed state is prepared in each of Alice and Bob's laboratories.
\item
Alice tosses a coin. If the coin lands on heads Alice performs a measurement $Q$ on the physical system in her laboratory, and if the coin lands on tails Alice performs a measurement $R$ on the physical system in her laboratory.
\item
Bob tosses a coin. If the coin lands on heads Bob performs a measurement $S$ on the physical system in his laboratory, and if the coin lands on tails Bob performs a measurement $T$ on the physical system in his laboratory.
\end{itemize}
It is also assumed that for each run of the experiment, the spacetime vector between the events corresponding to Alice's and Bob's measurements is spacelike, so that Alice's measurements are not in the future light cone of Bob's measurements, and vice versa. For those not familiar with spacetime geometry, this just means that if Alice sends a signal of light to Bob at the moment she performs her experiment, then the light signal will arrive in Bob's lab after he performs his experiment, and vice versa. Such a Bell scenario will be denoted by $(QR\,|\,ST)$.

Given a Bell scenario $(QR\,|\,ST)$, a physical theory should assign joint probability spaces $(\Omega_{QS},\mu_{QS})$, $(\Omega_{RS},\mu_{RS})$, $(\Omega_{QT},\mu_{QT})$ and $(\Omega_{RT},\mu_{RT})$ with each possible pair of Alice and Bob's measurements which may occur for each run of the Bell scenario. For example, $(\Omega_{QS},\mu_{QS})$ is the joint probability space corresponding to each run of the Bell scenario where Bob performs measurement $Q$ and Alice performs measurement $S$, so that $(\Omega_{QS},\mu_{QS})$ represents the probabilities of all possible outcomes of such a joint measurement scenario. Moreover, the marginals of these joint probability spaces should be the probability spaces $(\Omega_Q,\mu_Q)$, $(\Omega_R,\mu_R)$, $(\Omega_S,\mu_s)$ and $(\Omega_T,\mu_T)$ associated with Alice and Bob's individual measurements of $Q$, $R$, $S$ and $T$. 

We now formalize the notion of a physical theory which associates probability spaces with Bell scenarios, which we will refer to simply as a \emph{theory of measurement}. For this, we first set some notation in place.

\bnot
Given finite probability spaces $(\Omega_{Q},\mu_Q)$ and $(\Omega_{R},\mu_R)$, a joint probability space $(\Omega_Q\times \Omega_R,\mu_{QR})$ whose marginals are $(\Omega_Q,\mu_Q)$ and $(\Omega_R,\mu_R)$ will be denoted by $(\Omega_{QR},\mu_{QR})$, and the associated projection maps will be denoted by  
\[
(\Omega_{QR},\mu_{QR})\overset{\pi_Q}\lra (\Omega_Q,\mu_Q) \quad \text{and} \quad (\Omega_{QR},\mu_{QR})\overset{\pi_R}\lra (\Omega_R,\mu_R) \, .
\]
\enot

\bd
Let $(\Omega_{Q},\mu_Q)$, $(\Omega_{R},\mu_R)$, $(\Omega_{S},\mu_S)$ and $(\Omega_{T},\mu_T)$ be finite probability spaces. A \define{Bell square} is a diagram in $\bold{FinProb}$ of the following form:
\be \label{EPRXSXQ347}
\xy0;/r.25pc/:
(-35,0)*+{(\Omega_{RT},\mu_{RT})}="1";
(35,0)*+{(\Omega_{QS},\mu_{QS})}="2";
(-35,25)*+{(\Omega_{R},\mu_R)}="3";
(0,25)*+{(\Omega_{RS},\mu_{RS})}="4";
(35,25)*+{(\Omega_{S},\mu_{S})}="5";
(35,-25)*+{(\Omega_{Q},\mu_{Q})}="6";
(0,-25)*+{(\Omega_{QT},\mu_{QT})}="8";
(-35,-25)*+{(\Omega_{T},\mu_T)}="9";
{\ar"4";"3"_{\pi_{R}}};
{\ar"4";"5"^{\pi_{S}}};
{\ar"2";"5"_{\pi_{S}}};
{\ar"2";"6"^{\pi_{Q}}};
{\ar"8";"6"_{\pi_{Q}}};
{\ar"8";"9"^{\pi_{T}}};
{\ar"1";"9"_{\pi_{T}}};
{\ar"1";"3"^{\pi_{R}}};
\endxy 
\ee
\ed

\bd \label{TXMXS347}
A \define{theory of measurement} $\mathscr{M}$ associates each element in a collection of Bell scenarios $\{(QR\,|\, ST)\}$ with a unique Bell square $\mathscr{M}(QR\,|\,ST)\in \mathfrak{Diag}(\bold{FinProb})$. 
\ed

There is a theory of measurement $\mathscr{Q}$ referred to as \emph{quantum theory}, which is our best description of reality at the micro-scale. While quantum theory has a reputation of being difficult to understand, and has an ontological status which is highly controversial, in the context of Bell scenarios it may be thought of simply a theory of measurement as in Definition~\ref{TXMXS347} (which some might glibly label as the "Copenhagen interpretation"). In particular, for Bell scenarios where Alice and Bob are performing measurements on a quantum system, such as spin measurements on an electron, quantum theory yields a precise method for the assignment of Bell squares $\mathscr{Q}(QR\,|\,ST)$. The method prescribed by quantum theory involves associating a matrix $\rho_{AB}$ with a single run of a given Bell scenario, and then associating matrices with the measurements of $Q$, $R$, $S$ and $T$ in such a way that taking the trace of the associated matrices multiplied by $\rho_{AB}$ yields the associated probabilities\footnote{We note that much of the aforementioned controversy surrounding quantum theory stems from the lack of a general consensus regarding the assignment of an ontological meaning to the matrix $\rho_{AB}$.}. For those not at all familiar with the technical details of quantum theory, we will show how this is done for a specific Bell scenario in the appendix (we refer to the reader to Ref.~\cite{NiCh11} for further details). 

%%%%%%%%%%%%%%%%%%%%%%%%%%%%
\section{The principle of local realism} \label{RLX87}
%%%%%%%%%%%%%%%%%%%%%%%%%%%%

In this section, we use Bell squares to formulate a simple, mathematically precise definition of what it means for a theory of measurement (as in Definition~\ref{TXMXS347}) to satisfy a physical principle which is commonly referred to as \emph{local realism}. As its name suggests, the principle of local realism is the combination of two principles, namely, \emph{locality} and \emph{realism}. In the context of Bell scenarios, locality is taken to mean that Bob's measurement outcomes are unaffected in any way by the particular measurements Alice makes, and vice versa. Such a notion of locality is consistent with that of locality in special relativity, as we have assumed that neither Alice nor Bob are in the other's future light cone for each run of the Bell scenario. On the other hand, realism is taken to mean that the properties of Alice and Bob's physical systems which are revealed by measurements $Q$, $R$, $S$ and $T$, are all simultaneously well-defined for each run of the Bell scenario, irrespective of which measurements Alice and Bob actually perform. Assuming that a physical theory should satisfy realism is certainly consistent with our everyday experience of macroscopic random variables.  For example, if $Q$ and $R$ correspond to measurements of our weight and blood pressure, then it seems quite reasonable and intuitive that our weight and blood pressure both exist simultaneously, irrespective of which one we choose to measure.

 We now use Bell squares to formulate a precise definition of local realism in terms of single diagram in $\bold{FinProb}$. 

\bd \label{LCXRXSMX57}
A theory of measurement $\mathscr{M}$ is said to satisfy \define{local realism} if and only if for every Bell scenario $(QR\,|\,ST)$, there exists a common extension $(\Omega,\mu)$ of the joint probability spaces $(\Omega_{QS},\mu_{QS})$, $(\Omega_{RS},\mu_{RS})$, $(\Omega_{RT},\mu_{RT})$ and $(\Omega_{QT},\mu_{QT})$ appearing in the Bell square $\mathscr{M}(QS\,|\,ST)$ \eqref{EPRXSXQ347} such that the following diagram commutes.
\be \label{BTDGXS71}
\xy0;/r.25pc/:
(0,0)*+{(\Omega,\mu)}="0";
(-35,0)*+{(\Omega_{RT},\mu_{RT})}="1";
(35,0)*+{(\Omega_{QS},\mu_{QS})}="2";
(-35,30)*+{(\Omega_{R},\mu_R)}="3";
(0,30)*+{(\Omega_{RS},\mu_{RS})}="4";
(35,30)*+{(\Omega_{S},\mu_{S})}="5";
(35,-30)*+{(\Omega_{Q},\mu_{Q})}="6";
(-35,30)*+{(\Omega_{R},\mu_R)}="7";
(0,-30)*+{(\Omega_{QT},\mu_{QT})}="8";
(-35,-30)*+{(\Omega_{T},\mu_T)}="9";
{\ar"0";"1"_{\pi_{RT}}};
{\ar"0";"2"^{\pi_{QS}}};
{\ar"0";"8"_{\pi_{QT}}};
{\ar"0";"4"^{\pi_{RS}}};
{\ar"4";"3"_{\pi_{R}}};
{\ar"4";"5"^{\pi_{S}}};
{\ar"2";"5"_{\pi_{S}}};
{\ar"2";"6"^{\pi_{Q}}};
{\ar"8";"6"_{\pi_{Q}}};
{\ar"8";"9"^{\pi_{T}}};
{\ar"1";"9"_{\pi_{T}}};
{\ar"1";"7"^{\pi_{R}}};
\endxy
\ee
In such a case, $(\Omega,\mu)$ is said to be a \define{local realization} of the Bell square \eqref{EPRXSXQ347}. 
%A theory of measurement $\mathscr{M}$ which does not satisfy realism is then said to be \define{contextual}.
\ed

Suppose $\mathscr{M}$ is a theory of measurement which satisfies local realism, and suppose $(\Omega,\mu)$ is a local realization of the Bell square $\mathscr{M}(QR\,|\,ST)$ associated with a Bell scenario $(QR\,|\,ST)$, as in \eqref{BTDGXS71}. In such a case, $(\Omega,\mu)$ is a probability space encapsulating all possible outcomes of the Bell scenario $(QR\,|\,ST)$, so that an element $\omega\in \Omega$ is a global description of one of the possible outcomes of a single run of the Bell scenario $(QR\,|\,ST)$. Now suppose for a given outcome $\omega\in \Omega$ that Bob measures $S$. In such a case, the element $(\pi_S\circ \pi_{QS})(\omega)\in \Omega_S$ then corresponds to Bob's individual outcome when Alice measures $Q$, while the element $(\pi_S\circ \pi_{RS})(\omega)\in \Omega_S$ corresponds to Bob's individual outcome when Alice measures $R$. The commutivity of diagram \eqref{BTDGXS71} then ensures that
\be \label{CMXTXS67}
(\pi_S\circ \pi_{QS})(\omega)=(\pi_S\circ \pi_{RS})(\omega)\in \Omega_S\, ,
\ee
so that Bob's individual outcome when measuring $S$ is unaffected by whether Alice measures $Q$ or $R$. As the same exact argument applies for when Bob measures $T$, it follows that for every run of the Bell scenario $(QR\,|\,ST)$, Bob's measurements are unaffected by Alice's measurements. Moreover, as the above argument applies \emph{mutatis mutandis} to Alice's individual measurements in comparison to which measurement Bob makes, it follows that for every run of the Bell scenario $(QR\,|\,ST)$ Alice's measurements are unaffected by Bob's measurements. And this is precisely what it means for the principle of locality to hold in a given Bell scenario. 

As for the principle of realism, note that equation \eqref{CMXTXS67} also implies that a well-defined outcome for Bob's measurement $S$ exists for every possible outcome $\omega\in \Omega$ of the Bell scenario $(QR\,|\,ST)$. Similarly, the fact that the diagram \eqref{BTDGXS71} commutes yields the equations 
\[
(\pi_T\circ \pi_{QT})(\omega)=(\pi_T\circ \pi_{RT})(\omega)\in \Omega_T\,\, \forall \omega\in \Omega , \quad (\pi_Q\circ \pi_{QS})(\omega)=(\pi_Q\circ \pi_{QT})(\omega)\in \Omega_Q\,\, \forall \omega\in \Omega\, ,
\]
and
\[
(\pi_R\circ \pi_{RS})(\omega)=(\pi_R\circ \pi_{RT})(\omega)\in \Omega_R\,\, \forall \omega\in \Omega\, ,
\]
so that well-defined values of $T$, $Q$ and $R$ exist for every run of the Bell scenario $(QR\,|\,ST)$, \emph{irrespective} of which measurements Alice and Bob actually perform. And this is precisely the what it means for the principle of realism to hold in a given Bell scenario.  

We note that when formulated individually, the properties of locality and realism may be defined in terms of two separate commutative diagrams, as in Figure~1. In particular, for each of the fixed measurements Alice and Bob may perform, the locality diagram requires a common extension of the two joint probability spaces mapping to the sample space of the fixed measurement. For example, when Alice measures $Q$, locality requires a common extension $(\Omega_{Q}^{ST},\mu_{Q}^{ST})$ of the sample spaces $(\Omega_{QS},\mu_{QS})$ and $(\Omega_{QT},\mu_{QT})$ such that the following diagram is commutative.
\be \label{ALXISQ37}
\xy0;/r.25pc/:
(0,12.5)*+{(\Omega_{Q}^{ST},\mu_{Q}^{ST})}="0";
(30,12.5)*+{(\Omega_{QS},\mu_{QS})}="2";
(30,-12.5)*+{(\Omega_{Q},\mu_{Q})}="6";
(0,-12.5)*+{(\Omega_{QT},\mu_{QT})}="8";
{\ar"0";"2"^{\pi_{QS}}};
{\ar"0";"8"_{\pi_{QT}}};
{\ar"2";"6"^{\pi_{Q}}};
{\ar"8";"6"_{\pi_{Q}}};
\endxy
\ee
In such a case, $(\Omega_{Q}^{ST},\mu_{Q}^{ST})$ is viewed as the joint sample space associated with Alice and Bob's measurements, restricted to when Alice measures $Q$. The commutativity of diagram \eqref{ALXISQ37} then ensures that Alice's measurement of $Q$ is unaffected by which measurement Bob performs. The diagram for realism on the other hand only requires a single common extension $(\Omega,\mu)$ of the sample spaces $(\Omega_Q,\mu_Q)$, $(\Omega_R,\mu_R)$, $(\Omega_S,\mu_S)$, and $(\Omega_T,\mu_T)$, so that the outcomes of $Q$, $R$, $S$ and $T$ are all well-defined for each run of the Bell scenario, irrespective of which measurements Alice and Bob actually perform. Such an $(\Omega,\mu)$ will be referred to as a \define{realization} of the associated Bell square. Strikingly enough, it just so happens that \emph{both} principles of locality and realism may be simultaneously encapsulated by diagram \eqref{BTDGXS71}.

\begin{figure}
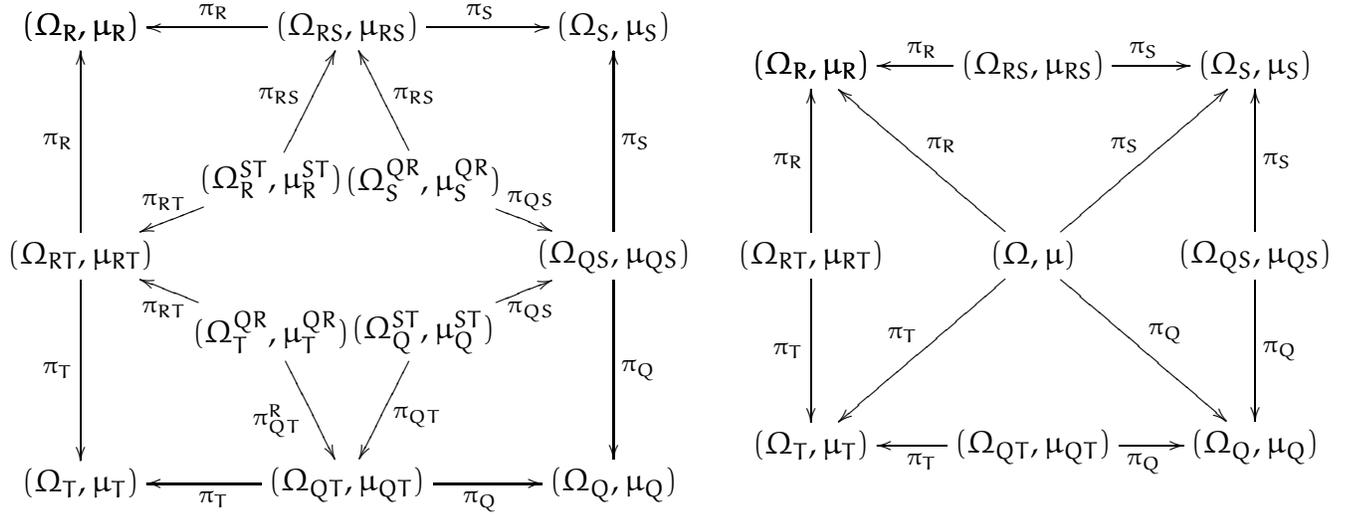
 \label{f1}
\begin{tabular}{cc}
$
\xy0;/r.24pc/:
(-10,10)*+{(\Omega_{R}^{ST},\mu_{R}^{ST})}="TL";
(10,10)*+{(\Omega_{S}^{QR},\mu_{S}^{QR})}="TR";
(10,-10)*+{(\Omega_{Q}^{ST},\mu_{Q}^{ST})}="BR";
(-10,-10)*+{(\Omega_{T}^{QR},\mu_{T}^{QR})}="BL";
(-35,0)*+{(\Omega_{RT},\mu_{RT})}="1";
(35,0)*+{(\Omega_{QS},\mu_{QS})}="2";
(-35,30)*+{(\Omega_{R},\mu_R)}="3";
(0,30)*+{(\Omega_{RS},\mu_{RS})}="4";
(35,30)*+{(\Omega_{S},\mu_{S})}="5";
(35,-30)*+{(\Omega_{Q},\mu_{Q})}="6";
(-35,30)*+{(\Omega_{R},\mu_R)}="7";
(0,-30)*+{(\Omega_{QT},\mu_{QT})}="8";
(-35,-30)*+{(\Omega_{T},\mu_T)}="9";
{\ar"TL";"1"_{\pi_{RT}}};
{\ar"TL";"4"^{\pi_{RS}}};
{\ar"TR";"4"_{\pi_{RS}}};
{\ar"TR";"2"^{\pi_{QS}}};
{\ar"BR";"2"_{\pi_{QS}}};
{\ar"BR";"8"^{\pi_{QT}}};
{\ar"BL";"1"^{\pi_{RT}}};
{\ar"BL";"8"_{\pi^{R}_{QT}}};
{\ar"4";"3"_{\pi_{R}}};
{\ar"4";"5"^{\pi_{S}}};
{\ar"2";"5"_{\pi_{S}}};
{\ar"2";"6"^{\pi_{Q}}};
{\ar"8";"6"_{\pi_{Q}}};
{\ar"8";"9"^{\pi_{T}}};
{\ar"1";"9"_{\pi_{T}}};
{\ar"1";"7"^{\pi_{R}}};
\endxy
$
&
$
\xy0;/r.20pc/:
(0,0)*+{(\Omega,\mu)}="0";
(-35,0)*+{(\Omega_{RT},\mu_{RT})}="1";
(35,0)*+{(\Omega_{QS},\mu_{QS})}="2";
(-35,30)*+{(\Omega_{R},\mu_R)}="3";
(0,30)*+{(\Omega_{RS},\mu_{RS})}="4";
(35,30)*+{(\Omega_{S},\mu_{S})}="5";
(35,-30)*+{(\Omega_{Q},\mu_{Q})}="6";
(-35,30)*+{(\Omega_{R},\mu_R)}="7";
(0,-30)*+{(\Omega_{QT},\mu_{QT})}="8";
(-35,-30)*+{(\Omega_{T},\mu_T)}="9";
{\ar"0";"3"_{\pi_{R}}};
{\ar"0";"6"^{\pi_{Q}}};
{\ar"0";"9"_{\pi_{T}}};
{\ar"0";"5"^{\pi_{S}}};
{\ar"4";"3"_{\pi_{R}}};
{\ar"4";"5"^{\pi_{S}}};
{\ar"2";"5"_{\pi_{S}}};
{\ar"2";"6"^{\pi_{Q}}};
{\ar"8";"6"_{\pi_{Q}}};
{\ar"8";"9"^{\pi_{T}}};
{\ar"1";"9"_{\pi_{T}}};
{\ar"1";"7"^{\pi_{R}}};
\endxy
$
\end{tabular}
\caption{(a) The diagrammatic formulation of locality alone assumes the existence of a common extension of each \emph{pair} of joint probability spaces mapping to a fixed corner of the Bell square, as in the diagram on the left. (b) The diagrammatic formulation of realism alone assumes the existence of a common extension of all four corners of the Bell square, as in the diagram on the right. Although these definitions are independent (neither implies the other), it turns out that they can be combined into a single diagrammatic formulation of local realism, as in \eqref{BTDGXS71}.}
\end{figure}

If $\mathscr{M}$ is a theory that satisfies realism and $(\Omega,\mu)$ is a realization of a Bell square $\mathscr{M}(QR\,|\,ST)$, then $(\Omega',\mu')$ is also a realization of $\mathscr{M}(QR\,|\,ST)$ for every extension $(\Omega'\mu')\to (\Omega,\mu)$ of the probability space $(\Omega,\mu)$. As such, a realization of a Bell square $\mathscr{M}(QR\,|\,ST)$ is never unique. While it may seem natural to fix a sample space associated with a random process, for example Terence Tao writes \cite{Tao}

"In practice, though, it is far more convenient to add new sources of randomness on the fly, if and when they are needed, and extend the sample space as necessary."

In the context of realistic descriptions of Bell scenarios, the probabilistic nature of Alice and Bob's measurements may always be viewed as due to unseen sources of randomness corresponding to an extension of the sample space $(\Omega,\mu)$ which is a realization of the associated Bell square, rather than any inherent randomness in their physical systems. For example, if $\Omega=\{H,T\}$ is the standard sample space associated with tossing a coin, then while it may appear as if an outcome of heads or tails is random, we know by Newton's Laws that the outcome of a coin toss is uniquely determined by the way in which the coin was tossed. As such, the probabilistic nature of a coin toss is due solely to our ignorance of the physical variables---such as torque and momentum---which deterministically yield the outcome of heads or tails. So in actuality, there exists an extension $(\Omega',\mu')\overset{f}\to (\Omega,\mu)$ of the form 
\[
\omega=f(x_1,x_2,\ldots,x_n)\in \Omega=\{H,T\}\, ,
\]
where the $x_i$ are the unseen physical variables associated with the coin toss. More generally, a realistic theory of measurement which postulates the existence of an extension $(\Omega',\mu')\overset{f}\to (\Omega,\mu)$ of the form $\omega=f(x_1,\ldots,x_n)$, where the variables $x_i$ are not accessible to observation or experiment is often referred to as a "hidden variables" theory. In a hidden variables theory, while the outcomes $\omega\in \Omega$ may appear to be random, such an appearance is due to our ignorance of the hidden variables $x_i$, similar to the case of a coin toss. 

We conclude this section with a proposition stating that locally realistic theories always exist.

\bn
Let $\mathscr{M}$ be a theory of measurement such that for every Bell scenario $(QR\,|\,ST)$, the joint distributions $\mu_{QS}$, $\mu_{RS}$, $\mu_{RT}$ and $\mu_{QT}$ appearing in the Ball square $\mathscr{M}(QR\,|\,ST)$ are the product distributions associated with $\mu_Q$, $\mu_R$, $\mu_S$ and $\mu_T$. Then $\mathscr{M}$ satisfies local realism.
\en

\bprf
The statement follows from the fact that given a Bell scenario $(QR\,|\,ST)$, the finite probability space $(\Omega_{QSRT},\mu_{Q}\times \mu_R\times \mu_S\times \mu_T)$ is a local realization of $\mathscr{M}(QR\,|\,ST)$.
\eprf

%%%%%%%%%%%%%%%%%%%%%%%%%%%%%%%%%%%%%%%
\section{The CHSH inequality and the failure of local realism}\label{CHSH}
%%%%%%%%%%%%%%%%%%%%%%%%%%%%%%%%%%%%%%%

Given a Bell scenario where Alice and Bob label their measurement outcomes by either $\pm 1$, one may use an assumption of local realism to derive an inequality bounding the statistics of Alice and Bob's joint measurements, which is referred to as the \emph{CHSH inequality} (after physicists Clauser, Holt, Shimony and Horne). As such, any theory of measurement which predicts a violation of the CHSH inequality cannot be locally real. In this section, we show that our diagrammatic formulation of local realism is consistent with these results, and we give an example showing how quantum theory yields probabilities which lead to a violation of the CHSH inequality. Before doing so however, we first introduce some notation.

\bnot
If $\X_Q\in \bold{RV}(\Omega_Q,\mu_Q)$ and $\X_R\in \bold{RV}(\Omega_R,\mu_R)$, and if $(\Omega_{QR},\mu_{QR})$ is a joint probability space associated with the $(\Omega_Q,\mu_Q)$ and $(\Omega_R,\mu_R)$, then we let $\X_{QR}\in \bold{RV}(\Omega_{QR},\mu_{QR})$ denote the product random variable given by
\[
\X_{QR}(q,r)=\X_Q(q)\X_R(r) \qquad \forall (q,r)\in \Omega_{QR}\, .
\]
\enot

\bt \label{TMX687}
Let $\mathscr{M}(QR\,|\,ST)$ be a Bell square associated with a theory of measurement $\mathscr{M}$ that satisfies local realism (as in Definition~\ref{LCXRXSMX57}), and let $\X_Q\in \bold{RV}(\Omega_Q,\mu_Q)$, $\X_R\in \bold{RV}(\Omega_R,\mu_R)$, $\X_S\in \bold{RV}(\Omega_S,\mu_S)$, and $\X_T\in \bold{RV}(\Omega_T,\mu_T)$ be $\{\pm 1\}$-valued random variables. Then
\be \label{BNXQLX67}
\bold{E}(\X_{QS})+\bold{E}(\X_{RS})+\bold{E}(\X_{RT})-\bold{E}(\X_{QT})\leq 2\, ,
\ee
where $\X_{QS}$, $\X_{RS}$, $\X_{RT}$ and $\X_{QT}$ are the product random variables associated with the joint probability spaces $(\Omega_{QS},\mu_{QS})$, $(\Omega_{RS},\mu_{RS})$, $(\Omega_{RT},\mu_{RT})$ and $(\Omega_{QT},\mu_{QT})$ appearing in the Bell square $\mathscr{M}(QR\,|\,ST)$.
\et

\bprf %[Proof of Theorem~\ref{TMX687}]
Since $\mathscr{M}$ satisfies local realism, there exists a local realization $(\Omega,\mu)$ of the Bell square $\mathscr{M}(QR\,|\,ST)$. Now let $\widetilde{\X}_{QS}, \widetilde{\X}_{RS}, \widetilde{\X}_{RT}, \widetilde{\X}_{QT}\in \bold{RV}(\Omega,\mu)$ be the pullbacks of $\X_{QS}$, $\X_{RS}$, $\X_{RT}$ and $\X_{QT}$ to $(\Omega,\mu)$, let $\X\in \bold{RV}(\Omega,\mu)$ be the random variable given by
\[
\X=\widetilde{\X}_{QS}+\widetilde{\X}_{RS}+ \widetilde{\X}_{RT}-\widetilde{\X}_{QT}\, ,
\]
and for each $\omega\in \Omega$, let
\begin{align*}
q^{S,\omega}&=(\pi_Q\circ \pi_{QS})(\omega)\, ,  &s^{Q,\omega}&=(\pi_S\circ \pi_{QS})(\omega)\, , \\
q^{T,\omega}&=(\pi_Q\circ \pi_{QT})(\omega)\, ,  &s^{R,\omega}&=(\pi_S\circ \pi_{RS})(\omega)\, , \\
r^{S,\omega}&=(\pi_R\circ \pi_{RS})(\omega)\, ,  &t^{Q,\omega}&=(\pi_T\circ \pi_{QT})(\omega)\, , \\
r^{T,\omega}&=(\pi_R\circ \pi_{RT})(\omega)\, ,  &t^{R,\omega}&=(\pi_T\circ \pi_{RT})(\omega)\, . \\
\end{align*}
For all $\omega\in \Omega$ we then have
\begin{align*}
\X(\omega)&=\widetilde{\X}_{QS}(\omega)+\widetilde{\X}_{RS}(\omega)+ \widetilde{\X}_{RT}(\omega)-\widetilde{\X}_{QT}(\omega) \\
&=\X_{QS}(\pi_{QS}(\omega))+\X_{RS}(\pi_{RS}(\omega))+ \X_{RT}(\pi_{RT}(\omega))-\X_{QT}(\pi_{QT}(\omega)) \\
&=\X_{QS}(q^{S,\omega},s^{Q,\omega})+\X_{RS}(r^{S,\omega},s^{R,\omega})+\X_{RT}(r^{T,\omega},t^{R,\omega})-\X_{QT}(q^{T,\omega},t^{Q,\omega}) \\
&=\X_{Q}(q^{S,\omega})\X_S(s^{Q,\omega})+\X_{R}(r^{S,\omega})\X_S(s^{R,\omega})+\X_{R}(r^{T,\omega})\X_T(t^{R,\omega})-\X_{Q}(q^{T,\omega})\X_T(t^{Q,\omega})\, .
\end{align*}
Now since the diagram \eqref{BTDGXS71} associated with the Bell square $\mathscr{M}(QR\,|\,ST)$ commutes, we have
\[
\pi_Q\circ \pi_{QS}=\pi_Q\circ \pi_{QT}\, , \quad  \pi_R\circ \pi_{RS}=\pi_R\circ \pi_{RT}\, , \quad  \pi_S\circ \pi_{QS}=\pi_S\circ \pi_{RS} \quad \text{and} \quad \pi_T\circ \pi_{RT}=\pi_T\circ \pi_{QT}\, ,
\]
from which it follows that for each $\omega\in \Omega$ we have
\[
q^{S,\omega}=q^{T,\omega}\, , \quad r^{S,\omega}=r^{T,\omega}\, , \quad s^{Q,\omega}=s^{R,\omega} \quad \text{and} \quad t^{Q,\omega}=t^{R,\omega}\, .
\]
It then follows that for all $\omega\in \Omega$ we have
\begin{align*}
\X(\omega)&=\X_{Q}(q^{S,\omega})\X_S(s^{Q,\omega})+\X_{R}(r^{S,\omega})\X_S(s^{R,\omega})+\X_{R}(r^{T,\omega})\X_T(t^{R,\omega})-\X_{Q}(q^{T,\omega})\X_T(t^{Q,\omega}) \\
&=\X_{Q}(q^{S,\omega})\X_S(s^{Q,\omega})+\X_{R}(r^{S,\omega})\X_S(s^{Q,\omega})+\X_{R}(r^{S,\omega})\X_T(t^{R,\omega})-\X_{Q}(q^{S,\omega})\X_T(t^{R,\omega}) \\
&=\Big(\X_{Q}(q^{S,\omega})+\X_{R}(r^{S,\omega})\Big)\X_S(s^{Q,\omega})+\Big(\X_{R}(r^{S,\omega})-\X_{Q}(q^{S,\omega})\Big)\X_T(t^{R,\omega}) \\
&\leq 2\, ,
\end{align*}
where the last inequality follows from the fact that $\X_Q$, $\X_R$, $\X_S$ and $\X_T$ are all $\{\pm 1\}$-valued random variables. It then follows that $\bold{E}(\X)\leq 2$ (since the expected value of a random variable is always bounded above by it maximum), thus setting $\aleph=\bold{E}(\X_{QS})+\bold{E}(\X_{RS})+\bold{E}(\X_{RT})-\bold{E}(\X_{QT})$, we have
\begin{align*}
\aleph&=\bold{E}(\X_{QS})+\bold{E}(\X_{RS})+\bold{E}(\X_{RT})-\bold{E}(\X_{QT}) && \text{by definition of $\aleph$} \\
&=\bold{E}(\widetilde{\X}_{QS})+\bold{E}(\widetilde{\X}_{RS})+\bold{E}(\widetilde{\X}_{RT})-\bold{E}(\widetilde{\X}_{QT}) && \text{by Proposition~\ref{EXPXT17XPB}} \\
&=\bold{E}\Big(\widetilde{\X}_{QS}+\widetilde{\X}_{RS}+\widetilde{\X}_{RT}-\widetilde{\X}_{QT}\Big) && \text{by linearity of expectation} \\
&=\bold{E}(\X) && \text{by definition of $\X$} \\
&\leq 2\, ,
\end{align*}
as desired.
\eprf

The following result then follows immediately from Theorem~\ref{TMX687}. 

\bc[A numerical test for the failure of local realism] \label{CRXSXT847}
Let $\mathscr{M}(QR\,|\,ST)$ be a Bell square associated with a theory of measurement $\mathscr{M}$, and suppose $\X_Q\in \bold{RV}(\Omega_Q,\mu_Q)$, $\X_R\in \bold{RV}(\Omega_R,\mu_R)$, $\X_S\in \bold{RV}(\Omega_S,\mu_S)$, and $\X_T\in \bold{RV}(\Omega_T,\mu_T)$ are $\{\pm 1\}$-valued random variables such that 
\[
\bold{E}(\X_{QS})+\bold{E}(\X_{RS})+\bold{E}(\X_{RT})-\bold{E}(\X_{QT})>2\, .
\]
Then $\mathscr{M}$ does not satisfy local realism.
\ec

The next example uses Corollary~\ref{CRXSXT847} to show that quantum theory does not satisfy local realism. As a consequence of this example, it follows that any theory of measurement which reproduces the probabilities of quantum theory cannot be locally real, a result which is known as Bell's Theorem \cite{Bell64}. 

\bx[Quantum theory is not locally real] \label{CNTXBSQX737}
As mentioned at the end end of Section~\ref{BSXTMXS47}, there is a theory of measurement referred to as \emph{quantum theory}, which we will denote by $\mathscr{Q}$. A typical Bell scenario considered in quantum theory is the following. At each run of the Bell scenario an electron is prepared in a certain state in each of Alice and Bob's laboratories. Given such an electron, one may subject it to a Stern-Gerlach apparatus situated in a certain direction $\vec{u}$, where $\vec{u}$ is a unit vector in $\R^3$. In the presence of the magnetic field induced by the Stern-Gerlach apparatus, the electron is deflected either in an upward direction or a downward direction. In the former case the electron is then said to be "spin up" with respect to the $\vec{u}$-direction, and in the latter case the electron is said to be "spin down" with respect to the $\vec{u}$-direction. The measurements $Q$, $R$, $S$ and $T$ of the associated Bell scenario then correspond to spin measurements in distinct directions $\vec{u}_{Q}$, $\vec{u}_{R}$, $\vec{u}_{S}$ and $\vec{u}_{T}$. It then follows that we may take the associated sample spaces $\Omega_Q$, $\Omega_R$, $\Omega_S$ and $\Omega_T$ to be 2-element sets, corresponding to whether a given electron happens to be spin up or spin down. It turns out that there is a specific Bell scenario $(QR\,|\,ST)$ of this type in which the states of the two electrons at each run of the Bell scenario are said to be \emph{entangled}, for which quantum theory associates the Bell square $\mathscr{Q}(QR\,|\,ST)$ determined by the following data\footnote{We will show how these probabilities are uniquely determined by quantum theory in the appendix.}.
\begin{align*}
\underline{(\Omega_{QS},\mu_{QS})}: && \underline{(\Omega_{RS},\mu_{RS})}:&\\
\mu_{QS}(q_1,s_1)&=\frac{1}{8-4\sqrt{2}}& \mu_{RS}(r_1,s_1)&= \frac{1}{8-4\sqrt{2}} \\
\mu_{QS}(q_1,s_2)&=\frac{1}{8+4\sqrt{2}}& \mu_{RS}(r_1,s_2)&=\frac{1}{8+4\sqrt{2}}   \\
\mu_{QS}(q_2,s_1)&=\frac{(1-\sqrt{2})^2}{8-4\sqrt{2}}& \mu_{RS}(r_2,s_1)&=\frac{3-2\sqrt{2}}{8-4\sqrt{2}}   \\
\mu_{QS}(q_2,s_2)&=\frac{(1+\sqrt{2})^2}{8+4\sqrt{2}}& \mu_{RS}(r_2,s_2)&=\frac{3+2\sqrt{2}}{8+4\sqrt{2}}   
\end{align*}
\begin{align*}
\underline{(\Omega_{RT},\mu_{RT})}: && \underline{(\Omega_{QT},\mu_{QT})}:&\\
\mu_{RT}(r_1,t_1)&=\frac{3+2\sqrt{2}}{8+4\sqrt{2}}& \mu_{QT}(q_1,t_1)&=\frac{1}{8+4\sqrt{2}}  \\
\mu_{RT}(r_1,t_2)&=\frac{3-2\sqrt{2}}{8-4\sqrt{2}}& \mu_{QT}(q_1,t_2)&=\frac{1}{8-4\sqrt{2}}  \\
\mu_{RT}(r_2,t_1)&=\frac{1}{8+4\sqrt{2}}& \mu_{QT}(q_2,t_1)&=\frac{(1+\sqrt{2})^2}{8+4\sqrt{2}}  \\
\mu_{RT}(r_2,t_2)&=\frac{1}{8-4\sqrt{2}}& \mu_{QT}(q_2,t_2)&=\frac{(1-\sqrt{2})^2}{8-4\sqrt{2}} 
\end{align*}

Now let $\X_Q\in \bold{RV}(\Omega_Q,\mu_Q)$, $\X_R\in \bold{RV}(\Omega_R,\mu_R)$, $\X_S\in \bold{RV}(\Omega_S,\mu_S)$ and $\X_T\in \bold{RV}(\Omega_T,\mu_T)$ be the random variables given by
\[
\X_Q(q_i)=\X_R(r_i)=\X_S(s_i)=\X_T(t_i)=(-1)^{i+1} \qquad \forall \, i\in \{1,2\}\, .
\] 
We then have
\[
\bold{E}(\X_{QS})=\sum_{i,j=1}^{2}\X_{QS}(q_i,s_j)\mu_{QS}(q_i,s_j)=\sum_{i,j=1}^{2}\X_{Q}(q_i)\X_{S}(s_j)\mu_{QS}(q_i,s_j)=\frac{\sqrt{2}}{2}\, ,
\]
and similarly we find
\[
\bold{E}(\X_{RS})=\bold{E}(\X_{RT})=-\bold{E}(\X_{QT})=\frac{\sqrt{2}}{2}\, .
\]
It then follows that
\[
\bold{E}(\X_{QS})+\bold{E}(\X_{RS})+\bold{E}(\X_{RT})-\bold{E}(\X_{QT})=2\sqrt{2}>2\, ,
\]
thus the quantum theory $\mathscr{Q}$ does not satisfy local realism by Corollary~\ref{CRXSXT847}.  
\ex

\appendix

%%%%%%%%%%%%%%%%%%%%%%%%%%%%%%%%%%%%%%%%%
\section{Computing probabilities associated with a quantum Bell scenario}
%%%%%%%%%%%%%%%%%%%%%%%%%%%%%%%%%%%%%%%%%

In this section, we show how quantum theory yields the probabilities from Example~\ref{CNTXBSQX737}, which considers a Bell scenario where Alice and Bob are performing certain spin measurements on an electron (for a more comprehensive introduction to quantum theory we reccomend~\cite{NiCh11}). The first thing to know, is that in quantum theory states of (finite-dimensional) physical systems are mathematically described by $d\times d$ hermitian matrices whose eigenvalues form a probability distribution, where $d$ is the number of possible outcomes of a given measurement. For example, when measuring the spin of an electron there are only two possible outcomes, namely, spin up or spin down. As such, the state of the electron with respect to its spin is described by a $2\times 2$ hermitian matrix whose eigenvalues are $p$ and $1-p$ for some $p\in [0,1]$, and conversely every $2\times 2$ hermitian matrix of this form describes a valid spin state of an electron.

As mentioned in Example~\ref{CNTXBSQX737}, spin measurements on an electron can be made with respect to any direction described by a unit vector $\vec{u}\in \R^3$. And similar to how the state of an electron is described by a $2\times 2$ hermitian matrix, a spin measurement with respect to a unit vector $\vec{u}\in \R^3$ is also described by a $2\times 2$ hermitian matrix. In particular, if $\vec{u}=(u_x,u_y,u_z)$, then the spin measurement of an electron with respect to the $\vec{u}$-direction is described by the hermitian matrix $\sigma_{\vec{u}}$ given by
\[
\sigma_{\vec{u}}=u_x\sigma_x+u_y\sigma_y+u_z\sigma_z\, ,
\]
where $\sigma_x$, $\sigma_y$ and $\sigma_z$ are the \emph{Pauli matrices}, which are given by
\[
\sigma_x=\left(
\begin{array}{cc}
0&1 \\
1&0 \\
\end{array}
\right)\, , \quad \sigma_y=\left(
\begin{array}{cc}
0&-i \\
i&0 \\
\end{array}
\right)\, , \quad \text{and} \quad
\sigma_z=\left(
\begin{array}{cc}
1&0\\
0&\matminus{1} \\
\end{array}
\right)\, .
\]

In the Bell scenario corresponding to Example~\ref{CNTXBSQX737}, Alice's measurements $Q$ and $R$ correspond to spin measurements with respect to the unit vectors $(0,0,1)$ and $(1,0,0)$ respectively, and so are described by the matrices $M_Q$ and $M_R$ given by
\[
 M_Q=\sigma_z \quad \text{and} \quad M_R=\sigma_x\, . 
 \]
On the other hand, Bob's measurements $S$ and $T$ correspond to spin measurements with respect to unit vectors in the direction of $(\matminus{1},0,\matminus{1})$ and $(\matminus{1},0,1)$ respectively, and so are described by the matrices $M_S$ and $M_T$ given by
\[
M_S=\frac{1}{\sqrt{2}}\left(
\begin{array}{cc}
\matminus{1}&\matminus{1}\\
\matminus{1}&1 \\
\end{array}
\right) \qquad \text{and} \qquad M_T=\frac{1}{\sqrt{2}}\left(
\begin{array}{cc}
1&\matminus{1}\\
\matminus{1}&\matminus{1} \\
\end{array}
\right)\, .
\]   
As the matrices $M_Q$, $M_R$, $M_S$ and $M_T$ are all hermitian and have eigenvalues $\pm 1$, it follows from the spectral theorem that
\[
M_Q=\Pi_Q^{+}-\Pi_Q^{-}\, , \quad M_R=\Pi_R^{+}-\Pi_R^{-}\, , \quad M_S=\Pi_S^{+}-\Pi_S^{-}\, , \quad \text{and} \quad M_T=\Pi_T^{+}-\Pi_T^{-}\, ,
\]
where $\Pi_Q^{\pm}$, $\Pi_R^{\pm}$, $\Pi_S^{\pm}$ and $\Pi_T^{\pm}$ are all idempotent, positive semi-definite matrices such that for all $X\in \{P,Q,R,S\}$,  
\[
\Pi_X^{+}\Pi_X^{-}=\Pi_X^{-}\Pi_X^{+}=0 \quad \text{and} \quad \Pi_X^{+}+\Pi_X^{-}=\mathds{1}\,,
\]
where $\mathds{1}$ denotes the $2\times 2$ identity matrix. 

Now for each run of the Bell scenario, the joint state of Alice and Bob's electrons are described by the $4\times 4$ hermitian matrix $\rho_{\text{EPR}}$ given by
\[
\rho_{\text{EPR}}=\frac{1}{2}\left(
\begin{array}{cccc}
0&0&0&0\\
0&1&\matminus{1}&0\\
0&\matminus{1}&1&0\\
0&0&0&0\\
\end{array}
\right)\, ,
\] 
and in such a case, the $2\times 2$ hermitian matrices representing Alice and Bob's individual electrons are both $\mathds{1}/2$ (which is obtained from $\rho_{\text{EPR}}$ by a matrix operation referred to as the partial trace). In the case that Alice measures $Q$ and Bob measures $S$, let $\Omega_Q=\{q_1,q_2\}$ and $\Omega_S=\{s_1,s_2\}$, where $x_1$ corresponds to spin up and $x_2$ corresponds to spin down for $x\in \{q,s\}$. According to the \emph{Born rule} of quantum theory, the associated joint probability space $(\Omega_{QS},\mu_{QS})$ is given by
\begin{align*}
\mu_{QS}(q_1,s_1)&=\Tr\Big[\rho_{\text{EPR}}(\Pi_Q^{+}\otimes \Pi_S^{+})\Big]=\frac{1}{8-4\sqrt{2}} \\
\mu_{QS}(q_1,s_2)&=\Tr\Big[\rho_{\text{EPR}}(\Pi_Q^{+}\otimes \Pi_S^{-})\Big]=\frac{1}{8+4\sqrt{2}} \\
\mu_{QS}(q_2,s_1)&=\Tr\Big[\rho_{\text{EPR}}(\Pi_Q^{-}\otimes \Pi_S^{+})\Big]=\frac{(1-\sqrt{2})^2}{8-4\sqrt{2}} \\
\mu_{QS}(q_2,s_2)&=\Tr\Big[\rho_{\text{EPR}}(\Pi_Q^{-}\otimes \Pi_S^{-})\Big]=\frac{(1+\sqrt{2})^2}{8+4\sqrt{2}} \, ,\\
\end{align*}
where $\otimes$ denotes the Kronecker product of matrices. The cases of the joint measurements $QT$, $RS$ and $RT$ then follow \emph{mutatis mutandis}.

%%%%%%%%BIBLIOGRAPHY%%%%%%%%%%%%
\addcontentsline{toc}{section}{\numberline{}Bibliography}
\bibliographystyle{plain}
\bibliography{XTX}

\begin{bibdiv}
\begin{biblist}

\bib{Aspect_1981}{article}{
      author={Aspect, Alain},
      author={Grangier, Philippe},
      author={Roger, Gerard},
       title={{Experimental Tests of Realistic Local Theories via Bell's
  Theorem}},
        date={1981},
     journal={Phys. Rev. Lett.},
      volume={47},
       pages={460\ndash 6443},
}

\bib{Bell64}{article}{
      author={Bell, J.~S.},
       title={{On the Einstein-Podolsky-Rosen paradox}},
        date={1964},
     journal={Physics Physique Fizika},
      volume={1},
       pages={195\ndash 200},
}

\bib{Bohm52}{article}{
      author={Bohm, David},
       title={{A Suggested interpretation of the quantum theory in terms of
  hidden variables. 1.}},
        date={1952},
     journal={Phys. Rev.},
      volume={85},
       pages={166\ndash 179},
}

\bib{Bouwmeester_1998}{article}{
      author={Bouwmeester, Dik},
      author={Pan, Jian-Wei},
      author={Daniell, Matthew},
      author={Weinfurter, Harald},
      author={Zeilinger, Anton},
       title={{Observation of three photon Greenberger-Horne-Zeilinger
  entanglement}},
        date={1999},
     journal={Phys. Rev. Lett.},
      volume={82},
       pages={1345\ndash 1349},
      eprint={quant-ph/9810035},
}

\bib{Clauser_1969}{article}{
      author={Clauser, John~F.},
      author={Horne, Michael~A.},
      author={Shimony, Abner},
      author={Holt, Richard~A.},
       title={{Proposed experiment to test local hidden variable theories}},
        date={1969},
     journal={Phys. Rev. Lett.},
      volume={23},
       pages={880\ndash 884},
}

\bib{Cl69}{article}{
      author={Clauser, John~F.},
      author={Horne, Michael~A.},
      author={Shimony, Abner},
      author={Holt, Richard~A.},
       title={{Proposed experiment to test local hidden variable theories}},
        date={1969},
     journal={Phys. Rev. Lett.},
      volume={23},
       pages={880\ndash 884},
}

\bib{deBroglie_1926}{book}{
      author={de~Broglie, Louis},
       title={{Waves and motions}},
   publisher={Gauthier-Villars et Cie \'Editeurs},
     address={Paris},
        date={1926},
      volume={1},
}

\bib{EPR}{article}{
      author={Einstein, Albert},
      author={Podolsky, Boris},
      author={Rosen, Nathan},
       title={Can quantum-mechanical description of physical reality be
  considered complete?},
        date={1935},
     journal={Phys. Rev.},
      volume={47},
       pages={777\ndash 780},
         url={https://link.aps.org/doi/10.1103/PhysRev.47.777},
}

\bib{Clauser_1972}{article}{
      author={Freedman, Stuart~J.},
      author={Clauser, John~F.},
       title={{Experimental Test of Local Hidden-Variable Theories}},
        date={1972},
     journal={Phys. Rev. Lett.},
      volume={28},
       pages={938\ndash 941},
}

\bib{Fr20}{article}{
      author={Fritz, Tobias},
       title={A synthetic approach to {M}arkov kernels, conditional
  independence and theorems on sufficient statistics},
        date={2020},
     journal={Adv. Math.},
      volume={370},
       pages={107239},
      eprint={1908.07021},
         url={https://doi.org/10.1016/j.aim.2020.107239},
}

\bib{Greenberger_1989}{article}{
      author={Greenberger, Daniel~M.},
      author={Horne, Michael~A.},
      author={Zeilinger, Anton},
       title={{Going Beyond Bell\textquoteright{}s Theorem}},
        date={1989},
     journal={Fundam. Theor. Phys.},
      volume={37},
       pages={69\ndash 72},
      eprint={0712.0921},
}

\bib{Ma98}{book}{
      author={Mac~Lane, Saunders},
       title={Categories for the working mathematician},
     edition={Second ed.},
      series={Graduate Texts in Mathematics},
   publisher={Springer-Verlag, New York},
        date={1998},
      volume={5},
        ISBN={0-387-98403-8},
}

\bib{NiCh11}{book}{
      author={Nielsen, Michael~A.},
      author={Chuang, Isaac~L.},
       title={Quantum computation and quantum information},
     edition={10th Anniversary},
   publisher={Cambridge University Press, Cambridge},
        date={2011},
        ISBN={0-521-63235-8; 0-521-63503-9},
         url={https://doi.org/10.1017/CBO9780511976667},
}

\bib{Peres88}{article}{
      author={Peres, Asher},
       title={Quantum limitaions on measurement of magnetic flux},
        date={1988Oct},
     journal={Phys. Rev. Lett.},
      volume={61},
       pages={2019\ndash 2021},
}

\bib{Plavala_2021}{article}{
      author={Pl\'avala, Martin},
       title={{General probabilistic theories: An introduction}},
        date={2023},
     journal={Phys. Rept.},
      volume={1033},
       pages={1\ndash 64},
      eprint={2103.07469},
}

\bib{Segal_47}{article}{
      author={Segal, I.E.},
       title={Postulates for general quantum mechanics},
        date={1947},
     journal={Ann. Math.},
      volume={48},
       pages={930\ndash 948},
         url={https://www.jstor.org/stable/1969387},
}

\bib{Tao}{misc}{
      author={Tao, Terrence},
       title={254a, notes 0: A review of probability theory},
        date={2010},
  url={https://terrytao.wordpress.com/2010/01/01/254a-notes-0-a-review-of-probability-theory/},
        note={Available at
  \url{https://terrytao.wordpress.com/2010/01/01/254a-notes-0-a-review-of-probability-theory/}},
}

\end{biblist}
\end{bibdiv}

\end{document}